\documentclass[12pt]{article}

\def\R{I\!\!R}

\def\lim{\mathop {\rm lim}}

\begin{document}

\begin{center} \bf \large SINGULAR HEAT AND WAVE EQUATIONS ON THE EUCLIDIEN SPACE $\R^n$ 
\end{center}
\begin{center} \bf Mohamed Vall Ould Moustapha\\
\end{center}

\begin{abstract}
In this paper we give the explicit formulas for the solution of the singular generalized  heat and wave equations on the Euclidian space $\R^n$:\\
\end{abstract}
 Math. Subj. Classification 2010 : 35JO5, 35JO8, 35K08.

\section{ Introduction} In this paper we discuss the explicit formulas for the solution of the following singular generalized  heat and wave equations on the Euclidian space $R^n$:\\
$$ \left \{\begin{array}{cc}\left(\frac{\partial}{\partial t}+\frac{k}{t}\right)
u(t,X)=\Delta u(t,X);(t,X)\in \R^\ast_+\times R^n\\ u(0,X)=f(X) ; f\in
C^\infty(\R)\end{array}
\right.\eqno(1.1) $$
$$ \left \{\begin{array}{cc}\left(\frac{\partial}{\partial t}+\frac{k}{t}\right)\left(\frac{\partial}{\partial t}-\frac{k}{t}\right)
w(t, X)=\Delta w(t, X); (t, X)\in \R^\ast_+\times \R\\ w(0,X)=0,  w_t(0, X)=g(X), g\in
C^\infty(\R)\end{array}
\right.\eqno(1.2 )$$
where
$$\Delta=\frac{\partial^{2}}{\partial X_{1}^{2}}+\frac{\partial^{2}}{\partial X ^{2}_{2}}+...+\frac{\partial^{2}}{\partial X^{2}_{n}}\eqno(1.3)$$
is the usual n-dimensional Euclidian Laplacian on $R^n$  and $k$ is a real number.\\
The mathematical interest in these equations, however, comes mainly from the fact that the time inverse potential $\frac{k}{t}$ (resp. the time inverse square $\frac{k(1-k)}{t^2}$)
is homogeneous of degree -1 (resp. -2) and therefore scales exactly the same as $\partial/\partial t$ (resp. $\partial^{2}/\partial t^{2}$).\\
An inconvenient of the time dependent potential is the absence of the relation between the semi-groups of the Schr\"odinger equation and the spectral properties of the operator.
The space inverse potential $k/x$ is called Coulomb potential and is widely studied in physical and mathematical literature$[1]$.\\ 
 The space inverse square potential $k(1-k)/x^2$ arises in several contexts, one of them is the
Schr\"odinger equation in non relativistic quantum mechanics (Reed and Simon $[7]$) .
For example, the Hamiltonian for a spinzero particle in Coulomb field
gives rise to a Schr\"odinger operator involving the space inverse square
potential (Case $[2]$).
The Cauchy problem for the wave equation with the space inverse square potential in
Euclidean space $\R^n$ is extensively studied (Cheeger and Taylor $[3]$), (Planchon et al$[6]$).
The cases considered
frequentely are  $k=0$, the equations in
$(1.1)$ and $(1.2)$ then turn into the classical heat and wave equation on the Euclidean spaces $\R^n$
and these equations appear in several branches of mathematics
and physics (Folland $[4], p.143, 171$).
Our main objective of this paper is to solve the Cauchy problems $(1.1)$ and $(1.2).$
\section{Singular heat equation}
{\bf Theorem 2.1} The generalized singular heat equation in $(1.1)$ has the following general solution\\
$\varphi(t,X)=A t^{-n/2}{}_1F_1\left(\frac {n}{2}-k, \frac{ n}{2}, \frac{|X|^2}{4 t}\right)+$\\ $$B t^{-n/2}\exp{\left(-\frac{|X|^{2}}{4t}\right)}U\left(k, \frac{n}{2},\frac{|X|^{2}}{4t}\right)\eqno(2.1)$$
with $A$ and $B$ are complex constants and ${}_1F_1\left(a, c, z\right)$ and $U(a, c, z)$ are the confluent hypergeometric functions of the first and the second kind given respectively by 
 ([5],p.263):\\
$${}_1F_1(a; c; z)=\sum_{k=0}^{}\frac{(a)_k}{(c)_k k!} z^k\ \ \ \ c\neq 0,-1,-2,..... \eqno(2.2) $$
$$U(a, c, z)=\frac{\pi}{\sin\pi c}\left[\frac{{}_1F_1(a; c; z)}{\Gamma(c)\Gamma(1+a-c)}-z^{1-c}\frac{{}_1F_1(a+1-c; 2-c; z)}{\Gamma(a)\Gamma(2-c)}\right]\eqno(2.3)$$
where as usual $(a)_n$ is the Pochhamer symbol defined by
$$
  (a)_n=\frac{\Gamma(a+n)}{\Gamma(a)}\eqno(2.4)
$$
and $\Gamma$ is the classical Euler function.\\
{\bf Proof} Using the geodesic polar coordinates centred at $X$, $Y=X+r\omega , r>0 ;\omega \in S^{n-1}$
with $S^{n-1}$ is the sphere of dimension $n-1$ , and setting $y=r^{2}$ in the
generalized singular heat equation in $(1.1)$ we obtain 
$$\left[4y(\partial^{2}/\partial y^{2})+2n(\partial/\partial y)\right]\Psi(t,y)=\left[(\partial/\partial t)+(k/t)\right]\Psi (t,y)\eqno(2.5)$$
by the change of function and the change of the variable below.\\
$$\Psi (t,y)=t^{-n/2}\Phi(t,y) ;  z=-y/4t \eqno(2.6)$$
the equation $(2.5)$ is transformed into the following confluent  hypergeometric equation
$$\left[z(d^{2}/d z^{2})
 +((n/2)-z)(d/dz)\right]\Phi(z)-((n/2)-k)\Phi(z)=0\eqno(2.7)$$
with parameters
$a=n/2-k; c=n/2$,($[5]$ p.$268$). An appropriate independent  solutions of this equation are: ($[5]$ p.$270$)
${}_1F_{1}\left(a,c,z\right)$ and $\exp(z) U(c-a,c,-z)$.
From the formulas $(2.5),(2.6)$ and $(2 .7)$ we conlude that the function $\varphi$ in $(2.1)$ is the general solution of the generalized 
singular heat equation in $(1.1)$.\\

{\bf Theorem 2.2} For $n\geq 2$ and  $ k\neq 0,-1,-2,...$, the Cauchy problem for the singular generalized heat equation $(1.1)$ has the unique solution given by
$$u(t,X)=\int_{R^{n}}H^{k}_{n}(t, X, Y)f(Y)dm(Y)\eqno(2.8)$$
where
$$H^k_n(t,X,Y)=\Gamma(k)(4\pi t)^{-n/2}\exp{\left(-\frac{|X-Y|^{2}}{4t}\right)}U\left(k, \frac{n}{2}, \frac{|X-Y|^{2}}{4t}\right)\eqno(2.9)$$
and $U(a, c, z)$ is the confluent hypergeometric function of the second kind given in $(2.3)$.\\
{\bf Proof }
In view of the proposition 2.1, to finish the proof of the theorem it remains to show
the limit condition in $(1.1)$, for this we recall the asymptotic behavior of the degenerate confluent hypergeometric function $ U(a,c,z)$, ($[5]$ p.$288-289$),\\
for $ z\rightarrow +\infty $\\
$$U(a,c,z)=z^{-a}+ O(|z|^{-a-1}) \eqno(2.10)$$
and for $z\longrightarrow 0$ 
$$U(a,c,z)=(1/\Gamma(a))\left[\log z+ \psi(a)-2\gamma \right]+O(|z\log z|) ,c=1 \eqno(2.11)$$
$$U(a,c,z)=(\Gamma(c-1)/\Gamma(a))z^{1-c}+ O(1) , 1<\Re c <2 \eqno(2.12)$$
$$U(a,c,z)=(\Gamma(c-1)/\Gamma(a))z^{1-c}+ O(|\log z|) , c=2 \eqno(2.13)$$
$$U(a,c,z)=(\Gamma(c-1)/\Gamma(a))z^{1-c}+ O(|z|^{\Re c-2}) , \Re c\geq 2, c\neq 2\eqno(2.14)$$
Using the geodesic polar coordinates centred at $X$, and by setting  $y=r^{2};z=y/4t$ in $(1.1)$ we get
$$u(t,X)=(\Gamma(k)/2\pi^{n/2})\int^{\infty}_{0}\exp(-z)U(k,n/2,z)z^{(n/2)-1}f_X^{\#}(\sqrt{4tz})dz\eqno(2.15)$$
with
$$f_X^{\#}(r)=\int_{S^{n-1}}f(X+r\omega)d\omega\eqno(2.16)$$
Taking the limit in $(2.15)$ in view of the formulas $(2.10)-(2.14)$ we can reverse the limit and the integral and we obtain
$$\lim_{t\longrightarrow 0}u(t, X)=c_n f_X^{\#}(0)\int^{\infty}_{0}\exp(-z)U(k, n/2, z)z^{(n/2)-1}dz\eqno(2.17)$$
and by the formula ($[1]$ p.$266$):
$$\int^{\infty}_{0}\exp(-z)U(a,c,z)z^{c-1}dz=-\exp(-z)z^{c}U(a, c+1, z)\eqno(2.18)$$
$$\lim_{t\rightarrow 0}u(t, X)= \left[-\frac{\Gamma(k)}{\pi^{n/2}} f_X^{\#}(0)\exp(z)U(k,n/2,z)z^{(n/2)}\right]^{\infty}_0\eqno(2.19)$$
using again the formulas $(2.10)$ and $(2.14)$ we have 
$$\lim_{t\rightarrow 0}w(t, X)= \frac{\Gamma(k)}{\pi^{n/2}} \frac{2\pi^{n/2}}{\Gamma(n/2)}\frac{\Gamma(n/2)}{\Gamma(k)}f_X^{\#}(0)=f(X)\eqno(2.20)$$
The unequeness is clear from the properties of the confluent hypergeometric equation
($[5]$ p.268-270).

\section{The generalized singular wave equation on $\R^n$}

{\bf Theorem 3.1} 
 The generalized singular wave equation in $(1.2)$ has the following general solution
$$w(t, X, Y)=A t^{1-n}{}_2F_1\left(\frac {n-k}{2}, \frac{ n-1+k}{2},\frac{n+1}{2}, 1-\frac{|X-Y|^2}{t^2}\right)+$$
$$B \left(t^2-|X-Y|^2\right)^{(1-n)/2}{}_2F_1\left(\frac {1-k}{2}, \frac{ k}{2},\frac{3-n}{2}, 1-\frac{|X-Y|^2}{t^2}\right)\eqno(3.1)$$
with $A$ and $B$ are complex constants and ${}_2F_1\left(a, b, c, z\right)$ is the Gauss hypergeometric function given by
$$
  F(a,b,c;z)=\sum_{n=0}^{\infty}\frac{(a)_n(b)_n}{(c)_n n!}z^n,
  \quad |z|<1,
\eqno(3.2)$$
{\bf Proof} Using the geodesic polar coordinates centred at $X$, $Y=X+r\omega , r>0 ;\omega \in S^{n-1}$, and setting $y=r^{2}$ and $x=t^2$ in the
generalized singular wave equation in $(1.2)$ we obtain \\
$\left[4y(\partial^{2}/\partial y^{2})+2n(\partial/\partial y)\right]\Psi(x,y)=$\\ 
$$\left[4x(\partial^{2}/\partial x^{2})+2(\partial/\partial x)+(k(1-k)/x)\right]\Psi(x,y)\eqno(3.3)$$
setting
$$\Psi(x,y)=x^{-(n-1)/2}\Phi(x,y);  z=y/x\eqno(3.4)$$
we obtain the following Gauss hypergeometric equation\\
$$z(1-z)\frac{d^{2}}{dz^{2}}\Phi(z)+[n/2-(n+1/2)z]\frac{d}{dz}\Phi(z)-(n-k)(n-1+k)/4\Phi(z)=0\eqno(3.5)$$
with parameters: $a=(n-k)/2; b=(n-1+k)/2; c=n/2$.\\
The hypergeometric equation $(3.5)$  has the following system of solutions 
$([5], p.42-43)$
$$\Phi_1(z)=F((n-k)/2,(n-1+k)/2;(n+1)/2,1-z)\eqno(3.6)$$
and
$$\Phi_2(z)=(1-z)^{(1-n)/2}F((1-k)/2,k/2;(3-n)/2,1-z)\eqno(3.7)$$
hence the following functions satisfy  the generalized singular wave equation in $(1.2)$\\
$\varphi_1^k(t, X, Y)= t^{1-n}\times$ $$F\left((n-k)/2,n-1+k)/2;(n+1)/2,1-\frac{|X-Y|^2}{t^2}\right)\eqno(3.8)$$
$\varphi_2^k(t, X, Y)=\left(t^2-|X-Y|^2\right)^{(1-n)/2}\times$\\ 
$$F\left((1-k)/2,k/2;(3-n)/2,1-\frac{|X-Y|^2}{t^2}\right)\eqno(3.9)$$
and the proof of the theorem  3.1 is finished.\\
In the remainder of this section we present several lemmas.\\
{\bf Lemma 3.2} 
For $X, Y\in \R^n$ and $t\in \R^+$ set\\
$W_2(t, X, Y)=$ $$c_2 \left(t^2-|X-Y|^2\right)^{-1/2}{}_2F_1\left(\frac {1-k}{2}, \frac{ k}{2},\frac{1}{2}, 1-\frac{|X-Y|^2}{t^2}\right)\eqno(3.10)$$
and
$$c_2=\frac{\Gamma(1+k/2)\Gamma(3-k)/2)}{\pi^{3/2}}\eqno(3.11)$$
for $n$ even $n\geq 4$ \\
$W_n(t, X, Y)=$ $$c_n \left(t^2-|X-Y|^2\right)^{(1-n)/2}{}_2F_1\left(\frac {1-k}{2}, \frac{ k}{2},\frac{3-n}{2}, 1-\frac{|X-Y|^2}{t^2}\right)\eqno(3.12)$$
and
$$c_n=\frac{2^{n/2-1}(n-3)!!\Gamma(n/2)}{((n-2)/2)!\pi^{(n-1)/2}}c_2\eqno(3.13)$$
then for  $$A_{x}^a=
( a|X-Y|^2)^{-1} x^{1-a}\frac{\partial}{\partial x} x^a=(a |X-Y|^2)^{-1}\left(x\frac{\partial}{\partial x}+a\right)\eqno(3.14)$$ the following formulas hold\\
i)
$$A^{(n-3)/2}_{t^2}W_n^k(t, X, Y)=W^{k}_{n+2}(t, X, Y)\eqno(3.15)$$
ii)
$$W_n^k(t, X, Y)=c_nA_{t^2}^{\frac{n-3}{2}} A_{t^2}^{\frac{n-5}{2}} ... A_{t^2}^{\frac{1}{2}}W_2^k(t, X, Y)\eqno(3.16)$$
iii)For $g\in C_0^\infty(\R^n)$ we have\\
 $A_{x}^{\frac{n-3}{2}} A_{x}^{\frac{n-5}{2}} ... A_{x}^{\frac{1}{2}}\left[x^{1/2}g(\sqrt{x z})\right]=$\\
$$(\frac{y}{2})^{(n-2)/2}\frac{(n-2)/2 !}{(n-3)!!} x^{1/2} g(\sqrt{x z})+x\sum^{(n-2)/2}_{i=1}b_{i}x^{(i-1)/2}z^{i/2}g^{(i)}(\sqrt{x z})\eqno(3.17)$$
with $b_i$ ; $i=1,2,...,(n-2)/2$ are real constants.\\
{\bf Proof }:
To show $i)$ we use the formula $([5]$ p.$41)$
$$\frac{d}{dz}z^{c-1}F(a,b;c,z)=(c-1)z^{c-2}F(a,b;c-1,z)\eqno(3.18)$$
$ii)$ comes from $i)$.\\
iii) we can demonstrate iii) by induction over even $n\geq 4$.\\
{\bf Lemma 3.3}
For $z\longrightarrow 0$ we have:\\
i)
$ F((1+k)/2,1-k/2;(n+1)/2,1-z)= $
$$\frac{\Gamma((n+1)/2)\Gamma((2-n)/2)}{\Gamma((1
+k)/2)\Gamma((2-k)/2)}z^{(n-2)/2} +O(1)\eqno(3.19)
$$
ii)
$F((1-k)/2,k/2;1/2,1-z)=$ $$\frac{\Gamma(1/2)}{\Gamma(k/2)\Gamma((1-k)/2))}[1+o(\log z))]\eqno(3.20)
$$
$k\neq 0,-2,-4,....$\\
{\bf Proof}
i) is easily seen from the formula $([5], p.47)$\\
$F(a, b, c, z)=\frac{\Gamma(c)\Gamma(c-a-b)}{\Gamma(c-b)\Gamma(c-a)}F(a, b, a+b-c+1, 1-z)+$
$$(1-z)^{c-a-b}\frac{\Gamma(c)\Gamma(a+b-c)}{\Gamma(b)\Gamma(a)}F(c-a, c-b, c-a-b+1, 1-z)\eqno(3.21)
$$
ii) is a consequence of the formula $([5]$ p.$44)$\\
$F(a,b;a+b,z)=\left(\Gamma(a+b)/\Gamma(a)\Gamma(b)\right)\sum^{\infty}_{n=0}\times$\\
$$((a)_{n}(b)_{n}/(n!)^{2})[2\psi(n+1)-\psi(a+n)-\psi(b+n)-\log (1-z)](1-z)^{n}\eqno(3.22)
$$
$\arg(1-z)<\pi$;$|1-z|<1$\\
\section{Cauchy problem for the singular wave equation on $\R^n$, $n$ odd}
{\bf Theorem 4.1} Suppose $n$ is odd and
$$W_n^k(t, X, Y)=C_n t^{1-n}{}_2F_1\left(\frac {n-k}{2}, \frac{ n-1+k}{2},\frac{n+1}{2}, 1-\frac{|X-Y|^2}{t^2}\right)\eqno(4.1)$$
with \\ $ C_n=\frac{\Gamma(n/2) k(k-1)}{2\pi^{n/2}(n-1)}\times$\\
$$\left[\frac{\Gamma((n-k)/2)\Gamma((n-1-k)/2)}{\Gamma((n-k)/2)\Gamma((n-1-k)/2)-\Gamma(n/2)\Gamma((n-1)/2)}\right]\eqno(4.2)$$
If $g\in C_0^\infty(\R^n)$, the function
$$ w(t, X)=\int_{|X-Y|< t} W_n^k(t, X, Y)g(Y)dY\eqno(4.3)$$
solves the Cauchy problem for the generalized singular wave equation$ (1.2)$\\
{\bf Proof}
In view of the theorem 3.1, we see that the kernel in $(4.1)$  satisfies the generalized singular wave equation in $(1.2)$ and hence 
the function $w(t, X)$ in $(4.3)$ satisfies the same equation. To complete the proof of the theorem 4.1 it remains to show the limit conditions.
Using the geodesic polar coordinates and setting $y=r^{2};x=t^{2},z=y/x$ in $(4.3)$ we have\\
$ w(t,X)=C_n\int^{1}_{0}\frac{1}{2} g_X^{\#}(t\sqrt{z})z^{(n-2)/2} \times $ $$x^{1/2}F((n-k)/2,(n-1+k)/2;(n+1)/2,1-z)dz\eqno (4.4)
$$
with $g_X^{\#}(r)$ is as in $(2.16)$.
By the formula $([5]  p.47)$.
$$F(a,b;c,z)=(1-z)^{c-a-b}F(c-a,c-b;c,z)\eqno(4.5)$$
we can write\\
$$ w(t,X)=C_n\frac{t}{2}\int^{1}_{0}F((1+k)/2,1-k/2;(n+1)/2,1-z)g_X^{\#}(t\sqrt{z})dz\eqno(4.6) 
$$
hence by taking the limit in $(4.6)$ using the formula $(3.19)$ we can reverse the integral and the limit to obtain
$$\lim_{t\rightarrow 0}w(t,X)=0.\eqno(4.7)
$$
For the second condition we derive the expression in $(4.6)$, using again $(3.19)$ we can derive under the integral sign to obtain\\
$\frac{\partial }{\partial t} w(t,X)=$\\ $$C_n\frac{1}{2}\int^{1}_{0}g_X^{\#}(t\sqrt{z})F((1+k)/2,1-k/2;(n+1)/2,1-z)dz
+tO(1)\eqno(4.8) 
$$
Hence by taking the limit of $(4.8)$ in view of $(3.19)$ we can reverse the limit and the integral to write
$$\lim_{t\rightarrow 0}\frac{\partial}{\partial t}w(t,X)=C_n\frac{1}{2}g_X^{\#}(0)\int^{1}_{0}F((1+k)/2,1-k/2;(n+1)/2,1-z)dz\eqno(4.9)
$$
$$\lim_{t\rightarrow 0}\frac{\partial}{\partial t}w(t,X)=C_n\frac{1}{2}g_X^{\#}(0)\int^{1}_{0}F((1+k)/2,1-k/2;(n+1)/2, z)dz\eqno(4.10)
$$
In view of the formula ($[5]$ p.$41$)
$$\frac{d}{dz} F(a,b;c,z)=\frac{a b}{c}F(a+1,b+1;c+1,z)\eqno(4.11)
$$
we obtain\\
$\lim_{t\rightarrow 0}\frac{\partial}{\partial t}w(t,X)=-C_n(1/2)g_X^{\#}(0)\frac{n-1}{k(k-1)}\times$ $$\left[F((k-1)/2,-k/2;(n-1)/2, z)\right]^{1}_{0}\eqno(4.12)
$$
that is\\
$\lim_{t\rightarrow 0}\frac{\partial}{\partial t}w(t,X)=-C_n(1/2)g_X^{\#}(0)\frac{n-1}{k(k-1)}\times$ $$\left[F((k-1)/2,-k/2;(n-1)/2, 1)-1\right]\eqno(4.13)
$$
And by the formula $(4.2)$  and ($[5]$ p.$40$):\\
$$F(a,b;c,1)=\left(\Gamma(c)\Gamma(c-a-b)/\Gamma(c-a)\Gamma(c-b)\right); \Re(a+b-c)<0\eqno(4.14)
$$
$c\neq 0,-1,-2,-3,...$ .\\
 we obtain
$\lim_{t\rightarrow 0}\frac{\partial}{\partial t}w(t,X)=g(X)$.\\
\\
\section{Cauchy problem for the singular wave equation on the Euclidien plane $\R^2$ }

{\bf Theorem 5.1} Suppose $n=2$ and 
 $$W_2^k(t, X, Y)=c_2\left(t^{2}-|X-Y|^{2}\right)^{\frac{-1}{2}} F\left(\frac{k}{2}, \frac{1-k}{2}, \frac{1}{2};1-\frac{|X-Y|^{2}}{t^{2}}\right)\eqno(5.1)$$
with
$$c_2=\frac{\Gamma(1+k/2)\Gamma(3-k)/2)}{\pi^{3/2}}\eqno(5.2)$$
If $g\in C^\infty_0(R^2)$, the function
$$ w(t, X)=\int_{|X-Y|< t} W_2^k(t, X, Y)g(Y)dY\eqno(5.3)$$
solves the Cauchy problem $(1.2)$.\\
{\bf Proof}
From the theorem 3.1 we see that the functions $w(t, X)$ in $(5.3 )$  satisfies the generalized singular wave equation in $(1.2)$ .\\  
 Now to show the limit conditions, by the geodesic polar coordinates and the change of variables
 $y=r^2; x=t^2, z=y/x$ in $(5.3)$, we have for
for $n=2$ \\
$$w(t,X)=c_2(t/2)\int^1_0(1-z)^{-1/2}F((1-k)/2, k/2, 1/2, 1-z)g_X^{\#}(t\sqrt{z})dz\eqno(5.4)
$$

By taking the limit in $(5.4)$  we can use the formula $(3.20)$ to reverse the limit and the integral and to obtain 
$$\lim_{t\rightarrow 0}w(t,X)=0\eqno(5.5)
$$
Now to show the second condition we derive the expression $(5.4)$ and in view of the formula $(3.20)$ we can derive under the integral sign to obtain\\
$\frac{\partial}{\partial t}w(t,X)=c_2\frac{1}{2} \int^{1}_{0}(1-z)^{-1/2}\times$\\ $$F((1-k)/2,k/2;1/2,1-z)g_X^{\#}(t\sqrt{z})dz+t O(1)) g_X^{\#}(t\sqrt{z})dz\eqno(5.6)
$$
Using again the formula $(3.20)$ we can reverse the limit and the integral and we have
$$\lim_{t\longrightarrow 0}\frac{\partial}{\partial t} w(t,X)=c_2 \frac{1}{2}g_X^{\#}(0)\int^{1}_{0}(1-z)^{-1/2}F((1-k)/2,k/2;1/2,1-z)dz\eqno(5.7)
$$
$$\lim_{t\longrightarrow 0}\frac{\partial}{\partial t} w(t,X)=c_2 \frac{1}{2}g_X^{\#}(0) \int^{1}_{0}z^{-1/2}F((1-k)/2,k/2;1/2,z)dz\eqno(5.8)
$$
 we have by the formula $(3.18)$ 
$$\lim_{t\longrightarrow 0}\frac{\partial}{\partial t}w(t,X)=c_2\frac{1}{2}g_X^{\#}(0) 2z^{1/2}\left[F((1-k)/2,k/2;3/2,z)\right]_0^1 \eqno(5.9)
$$
and from $(4.14)$
$$\lim_{t\rightarrow 0}\frac{\partial}{\partial t}w(t, X)=c_2\pi^{3/2} / \Gamma((3-k)/2))\Gamma((1+k/2)g(X)=g(X)\eqno(5.10)
$$

\section{Cauchy problem for the singular wave equation on $\R^n$, $n\geq 4 $ even }
{\bf Theorem 6.1} Suppose $n$ is even and  $n\geq 4$,
let $W_2^k(t, X, Y)$ is as in theorem 5.1
and $A^a_{x}$ is as in $(3.11)$  and
$$c_n=\frac{(n-3)!!\Gamma(n/2)}{2^{1-n/2}((n-2)/2)!\pi^{(n-1)/2}}\eqno(6.1)$$
If $g\in C^\infty_0(R^n)$, the function
$$ w(t, X)=c_nA_{t^2}^{\frac{n-3}{2}} A_{t^2}^{\frac{n-5}{2}} ... A_{t^2}^{\frac{1}{2}}\int_{|X-Y|< t} W_2^k(t, X, Y)g(Y)dY\eqno(6.2)$$
solves the Cauchy problem $(1.2)$.\\
{\bf Proof}
In view of the theorem 3.1, we see that the functions $w(t, X)$ in $(6.2)$ satisfies the generalized singular wave equation in $(1.2)$ .\\  
To finish the proof of the theorem we show the limit condition in the even case $n\geq 4$:
using the geodesic polar coordinates and setting $y=r^{2}; x=t^{2};  z=y/x$ in $(6.2)$; we have:\\
for $n$ even $n\geq 4$:\\

$$w(t, X)=c_nc_2B_t\left[(t/2)\int^{1}_{0}(1-z)^{-1/2}
F((1-k)/2,k/2;1/2,1-z)g_X^{\#}(t\sqrt{z})dz\right]\eqno(6.3)$$
with
$$B_t=A_{t^2}^{\frac{n-3}{2}} A_{t^2}^{\frac{n-5}{2}} ... A_{t^2}^{\frac{1}{2}}$$
Using the formula iii) of lemma 3.2
we have\\
$w(t, X)=C_n\frac{t}{2} \int^{1}_{0}(1-z)^{-1/2}F((1-k)/2, k/2;1/2, 1-z)g_X^{\#}(t\sqrt{z})dz+$\\
$$t^2 \sum_{i=0}^{(n-2)/2}b_{i}t^{i-1}\int_0^{1}(1-z)^{-1/2}F((1-k)/2, k/2;1/2,1-z) z^{i/2}\widetilde{g}_X^{(i)}(t\sqrt{z})dz \eqno(6.4)$$
with
$$C_n=c_n c_22^{-(n-2)/2}\frac{(n-2)/2!}{(n-3)!!} $$
Taking the limit of the expression $(6.3)$
and using the formula $(3.20)$
we can reverse the integral and the limit  to obtain
$$\lim_{t\rightarrow 0}w(t,X)=0\eqno(6.5)$$
For the second condition  we derive the expression $(6.4)$ and by $(3.17)$ we can derive under the integral sign\\
$\frac{\partial}{\partial t}w(t,X)=c_n c_22^{-(n-2)/2}\frac{(n-2)/2!}{(n-3)!!}  \frac{1}{2} \int^{1}_{0}(1-z)^{-1/2}\times$
$$F((1-k)/2,k/2;1/2,1-z)g_X^{\#}(t\sqrt{z})dz
+t O(1)) g_X^{\#}(t\sqrt{z})dz\eqno(6.6)$$
Taking now the limit in $(6.6)$ and using again $(3.20)$ we can reverse the limit and the integral sign\\
$\lim_{t\longrightarrow 0}\frac{\partial}{\partial t}w(t,X)==c_n c_22^{-(n-2)/2}\frac{(n-2)/2!}{(n-3)!!}  \frac{1}{2} g_X^{\#}(0)\times$\\ $$\int^{1}_{0}(1-z)^{-1/2}F((1-k)/2, k/2;1/2, 1-z)dz\eqno(6.7)$$
$\lim_{t\longrightarrow 0}\frac{\partial}{\partial t}w(t,X)=$ $$=c_n 2^{-(n-2)/2}\frac{(n-2)/2!}{(n-3)!!}  \frac{1}{2} g_X^{\#}(0) \int^{1}_{0}z^{-1/2}F((1-k)/2,k/2;1/2,z)dz\eqno(6.8)$$
by the formula $(3.18)$ we have\\
$\lim_{t\longrightarrow 0}\frac{\partial}{\partial t}w(t, X)==c_n c_22^{-(n-2)/2}\frac{(n-2)/2!}{(n-3)!!}  \frac{1}{2} \times$ $$\widetilde{g}_X(0) 2z^{1/2}\left(F((1-k)/2,k/2;3/2,z)\right)_0^1 \eqno(6.9)$$
using the formula $(4.14)$ we have\\
$\lim_{t\rightarrow 0}\frac{\partial}{\partial t}w(t, X)
=c_{n }c_2 2^{-(n-2)/2}\frac{(n-2)/2!}{(n-3)!!}\times$ 
$$\frac{ \pi^{(n+2)/2}}{\Gamma((3-k)/2))\Gamma((1+k/2)\Gamma(n/2)}g(X)=g(X)\eqno(6.10)$$
\section{Applications}
{\bf Remark 7.1}: we have
$$\lim_{k\rightarrow 0}\left[\Gamma (k)\right]^{-1}H^{k}_{n}(t, X,Y)=K_{n}(t, X, Y)\eqno(7.1)$$
where
$$K_{n}(t, X,Y)=(4\pi t)^{-n/2}\exp{\left(-|X-Y|^{2}/4t\right)}\eqno(7.2)$$
is the classical heat kernel on $R^{n}$.\\
{\bf Corollary 7.2} The generalized Cauchy problem for the heat equation on $\R^n$:

$$ \left \{\begin{array}{cc}\left(\frac{\partial}{\partial t}\right)v(t, X)=0&;(t, X)\in \R^\ast_+\times \R^n\\  \lim t^{-k}v(t, X)=v_0(X) ~~; v_0\in C^\infty_0(\R^n)\end{array}
\right.\eqno(7.3) $$
has the unique solution given by 
$$v(t, X)=\int_{\R^n}K^k_n(t, X, Y)f(Y)dm(Y)\eqno(7.4)$$
where
$$K^k_n(t, X, Y)=\Gamma(k)t^k(4\pi t)^{-n/2}\exp{\left(-\frac{|X-Y|^2}{4t}\right)}U\left(k, \frac{n}{2}, \frac{|X-Y|^2}{4t}\right)\eqno(7.5)$$
{\bf Proof} The proof of this corollary is simple and is omitted.\\
{\bf Corollary 7.3} We have
$$\lim_{k\longrightarrow 0}W_n^k(t, X, Y)=W_n(t, X, Y)\eqno(7.6)$$
with $$W_n(t, X, Y)=(2\pi)^{-n/2}\left(t^2-|X-Y|^2\right)^{(1-n)/2}\eqno(7.7)$$
is the classical wave kernel on $\R^n$ $[4]$ \\ \\
\noindent
{\bf Proof} The proof of this corollary is simple and is left to the reader.

\begin{flushleft}
Al Jouf University\\
College of Sciences and Arts\\
Al-Qurayyat
Saoudi Arabia.\\
and\\
Universit\'e de Nouakchott Al-asriya\\
Facult\'e des Sciences et Techniques\\
Unit\'e de Recherche: {\bf \it Analyse, EDP et Mod\'elisation: (AEDPM)}\\
B.P: 5026, Nouakchott-Mauritanie\\
E-mail adresse:mohamedvall.ouldmoustapha230@gmail.com\\
\end{flushleft}

\end{document}